\documentclass{kluwer}    % Specifies the document style.

\newdisplay{guess}{Conjecture}

\begin{document}                                                                                   
\begin{article}
\begin{opening}         
\title{Microquasars: Open Questions and Future Perspectives} 
\author{I.F. \surname{Mirabel}}  
\runningauthor{F\'elix Mirabel}
\runningtitle{Prospects for the future}
\institute{CEA/DSM/SAp. CE-Saclay, France \& IAFE/CONICET, Argentina.}
\date{November 12, 2000}

\begin{abstract}
The discovery and subsequent study of microquasars lead to major 
progress in our understanding of: 
1) the nature of relativistic jets seen elsewhere in the universe, 
and  2) the connection between the accretion onto compact objects 
and the formation of collimated jets. A detailed 
account of the major progress accomplished until present  
was published in Annual Review of Astronomy \& Astrophysics 
(Mirabel \& Rodr\'\i guez, 1999). Here I review  
the questions that remain unanswered, as well as the 
future perspectives that this new field of research is opening.  
\end{abstract}

\end{opening}           

\section{Major progress accomplished}
\vskip .05in

In relativistic sources located in the Milky Way,   
two-sided moving jets can be observed, and therefore can be overcome 
several of the ambiguities that had 
dominated the physical interpretation of extraglactic jets,  
where so far only the motions of one-side of the jets 
could be followed. From the observation of two-sided 
moving jets in microquasars, the system of equations can be solved. 
For the first time, 
an upper limit of the distance to a microquasar was 
derived from the proper motions using special relativity constraints  
(see Mirabel \& Rodr\'\i guez 1999). 
\vskip .05in

Major progress has also been made in the understanding of the 
accretion/ejection phenomenology. From multiwavelength observations, 
the connection between accretion flow instabilities -observed in the X-rays-  
with the ejection of relativistic plasma -observed at radio, infrared, and 
possibly optical and X-ray wavelengths-, has been established on a firm basis. 
After the softer component of the X-ray emitting plasma disappears, the inner accretion disk is rapidly 
re-established and the plasma that produces the 
hard X-ray component, is blown away 
(Mirabel et al. 1998) in the form of collimated relativistic jets 
(Dhawan, Mirabel \& Rodr\'\i guez, 2000). 

\vskip .05in
The X-ray source GRS 1915+105 (Castro-Tirado et al. 1994) has become a main 
target to study accreting black holes of stellar mass. However,  
the discovery and study of other microquasars is important not only 
to enhance the statistical sample, but also to make progress 
on the different aspects of the diverse phenomenology  
that each new object of this class has revealed 
(Mirabel \& Rodr\'\i guez 1999).

\section{Open questions}  
                    % Produces section heading.  Lower-level
                    % sections are begun with similar 
                    % \subsection and \subsubsection commands.

{\bf What is a microquasar ?} 

By this term we usually design 
stellar-mass black holes that mimic, 
on a smaller scale, many of the phenomena seen in quasars. The question 
is whether under this concept we should also include neutron star 
systems with jets. Although the physical meaning of the 
quasar-microquasar analogy as proposed by Mirabel \& Rodr\'\i guez (1998) 
is valid for black holes, because in many cases the nature of 
the compact object is unknown, 
we may currently include in this subclass of stellar sources, neutron 
star binaries that produce relativistic jets.    
\vskip .05in

{\bf Are all accreting black holes of stellar mass microquasars ?} 

All microquasars are accreting compact objects of stellar mass, 
but is the reverse true ? 
This question can be re-formulated as follows: do all X-ray black hole 
binaries produce jets ? From the theoretical 
point of view jets are needed to liberate angular 
momentum. On the other hand, observations show that X-ray emitting 
black holes, in addition to possible synchrotron emission associated to 
sporadic outbursts, always exhibit flat-spectrum compact counterparts of 
synchrotron emission. It has been proposed (Fender, 2000) and demonstrated by 
high resolution images (Dhawan, Mirabel \& Rodr\'\i guez, 2000), that 
these flat spectrum compact radio counterparts are thick collimated jets 
of AU size scales. Furthermore, 
from multiwavelength observations at radio, infrared, and X-ray wavelengths 
of the large-scale jets in SS433/W50, we know   
that in some circumstances jets exist without being seen. Therefore, 
the answer to this question is that probably, all accreting black holes of 
stellar mass are microquasars. 

\vskip .05in

{\bf Are the Lorentz factors of the bulk motions in the jets from 
quasars different from those in microquasars ?}

In other words, 
is the quasar-microquasar analogy valid ? The microquasar jets observed 
so far have bulk speeds that are 
statistically smaler than in quasars. For instance, in quasars are 
found jets with bulk speeds of up to $\sim$20c, but no microquasar with 
such jets has so far been found. Is this due to some fundamental 
difference that would invalidate the physical analogy between quasars and 
microquasars, or is it rather due to selection effects related to 
the Doppler favoritism needed in order to measure the proper motion of 
the ejecta in distant quasars ? Although from the observational point of view 
this remains an open question, theoretically it is not clear why 
the bulk motion in jets from supermassive black holes would be different 
from that in the jets from stellar mass black holes.

\vskip .05in

{\bf Could the terminal velocity of the jet be a diagnostic 
for the nature of the collapsed object (NS versus BH) ?} 

In all areas of astrophysics there have been increasing evidences that 
accretion is always related to the production 
of collimated jets. The observed outflow speeds seem to be of the order of the 
escape velocity from the surface of the accreting object. Objects 
as diverse as very young stars, nuclei of planetary nebulae, and 
accreting white dwarfs have jets  with non-relativistic velocities 
($\sim$100-10000 km s$^{-1}$), whereas neutron stars and black holes 
produce jets with relativistic speeds V $\geq$ 0.1c (Livio, 1999; Mirabel 
\& Rodr\'\i guez, 1999). If the present 
trends between the escape velocities from the accreting objects and the 
velocity of the outflows were confirmed as a strong correlation, it would 
imply that gravity is important and dominates over MHD mechanisms (Meier 
et al. 1997). However, it is not clear whether we will be able to 
{\it discriminate between neutron 
stars and black holes by knowing the terminal velocity of the collimated 
outflows}. For instance, Sco X-1 which problably contains a neutron star, 
has lobes moving with an average velocity of 0.45c (Fomalont, Geldzahler \& 
Bradshaw, 2000). Since only a handful of relativistic 
jet sources have been discovered so far in the Galaxy (Mirabel \& 
Rodr\'\i guez, 1999), we need the discovery and study of more sources 
to obtain a statistically significant number to answer this question. 
  
\vskip .05in

{\bf Why are QPOs in microquasars and not in AGNs ?} 

The scale of time of the instabilities in the accretion disk of black 
holes is proportional to the mass of the black hole. QPOs with periods 
of $\sim$1 sec in black holes of 10 M$_{\odot}$ would 
correspond to QPOs of $\sim$1 day to $\sim$3 years in black holes 
with masses of 10$^6$ and 10$^9$ M$_{\odot}$ respectivelly. QPOs of 
1 sec in microquasars correspond to typical fluctuations of $\sim$5\% of 
the X-ray flux. At X-rays the companion binary stars are much dimmer 
than the accretion disk. On the other hand, the accretion 
disk in AGNs is cooler (Mirabel \& Rodr\'\i guez, 1999), and 
QPOs should be observed in the UV and optical, wavelengths 
at which stars mostly radiate, and interstellar absorption is important. 
Since AGNs may be blurred by nuclear starbursts, 
it may be difficult to detect fluctuations of $\sim$5\% from AGNs 
that are embedded in starburst nuclei. 

One then may ask about the analogous of the large scale oscillations of the 
flux ($\geq$ 50\%) with periods of $\sim$30 min regularly observed in 
GRS 1915+105 (Greiner, Morgan \& Remillard, 1996). In black holes 
with masses in the range of 10$^6$ to 10$^9$ M$_{\odot}$  analogous QPOs would 
correspond to periods in the range of 5 to 500 years, which are 
difficult to monitor in human time-scales (Sams, Eckart \& Sunyaev, 1996).    
 
\vskip .05in

{\bf To image the jet close to the black hole, do microquasars offer 
an advantage with respect to AGNs ?}

No. The Schwarzschild radii of BHs are proportional to their mass 
and in these units nearby AGNs can be imaged 
closer to the BH than microquasars. For instance, M87 which is in the 
Virgo cluster at a distance of 15 Mpc and contains a BH of 
3 10$^9$ M$_{\odot}$ has been imaged with a resolution of $\sim$50 times 
the Schwarzschild radius (Junor, Biretta \& Livio, 1999).  
A given angular size projected on the supermassive BH of M87 is in terms 
of its Schwarzschild radius 10$^4$ smaller than the same distance projected 
on a BH of 10 M$_{\odot}$ at a distance of 2 kpc from the Sun. 
Therefore, relative to supermassive BHs in the Local Universe, the 
study of stellar mass BHs presents an advantage because of the time scales, 
but not as far as the dimensions of length are concerned. 
   
\vskip .05in
{\bf What is the connection between accretion and ejection ?}

In GRS 1915+105 large amounts of X-ray emmiting plasma 
($\sim$10$^6$ L$_{\odot}$) disappear in less than a 
few seconds and soon after synchrotron jets are formed 
(Fender \& Pooley, 1998; Eikenberry et al. 1998; Mirabel et al. 1998). 
Recent analysis shows that the X-ray flux that suddenly disappears 
only corresponds to the softer component (2-20 keV). The formation of the 
jets starts later, at the time of a spike that consists of a 
sudden increase in the flux of the soft component simultaneous with a 
decrease of the flux of the harder component (20-60 keV). 
In the context of current models these observations imply that the 
inner accretion disk first disappears (Belloni et al. 1997), and when  
the inner accretion disk is being re-established, a shock is produced which 
triggers the blow up of the plasma that was emmiting the hard X-rays   
in the form of collimated jets at relativistic speeds (Dhawan, Mirabel 
\& Rodr\'\i guez, 1999). In GRS 1915+105, an analogous correlation between 
the sudden disappearance of the 20-100 keV measured by BATSE and major 
ejection events seen in the radio with the VLA had been observed (see Mirabel 
\& Rodr\'\i guez, 1999 for references and the discussion of this issue).     

\vskip .05in
{\bf What fraction of the inflow goes into the jets ?}

The observations described above indicate that the matter and energy 
emitting in the soft X-rays that suddenly disappears is perhaps 
advected into the BH, 
since it is not immediately translated neither in hard X-rays
nor in synchrotron jets. Because of the uncertainties in estimating 
the mass of the X-ray emitting plasma it is difficult to estimate 
the fraction of the inflow that goes into the jets. On the contrary, 
assuming equipartition, the amount of mass and energy of the jets can be 
calculated (Fender \& Pooley, 1998; Mirabel et al. 1998). 

Now there are increasing evidences that the power of the jets may be a 
large fraction of the accretion power. For instance, in its very high state, 
GRS 1915+105 may have a short-term jet power of $\sim$10$^{39}$ erg 
s$^{-1}$ (Mirabel \& Rodr\'\i guez, 1999; Fender, 2000), which is a large 
fraction of the observed accretion power. This is consistent with the model 
for a rapidly rotating BH with a high accretion rate by Meier (2000). 
\vskip .05in
{\bf What is the mechanism that launches the jets ?}

MHD power which was nicely reviewed by Meier (2001) and may play an 
important role. However, it is still unclear the relative importance 
of the rotating thin accretion disk mechanism by Blandford \& 
Payne (1982) and that from the frame-dragged accreting matter inside 
the ergosphere of a rotating BH (Blandford \& Znajek, 1977). 
The X-ray spectral properties of microquasars with powerful jets (e.g. 
GRS 1915+105, GRO J1655-40) indicate that these are rotating BHs with 
spins near maximum (Zhang, Cui \& Chen, 1997), which is consistent with the 
Blandford \& Znajek 
mechanism. On the other hand, the Blandford \& Payne mechanism is 
consistent with the following two recent observations: 1) the large 
opening angle of $\sim$60$^{\circ}$ of the jet in M87 
(Junor, Biretta \& Livio, 1999) which seems to be the 
two dimensional image of a jet with a magnetic polar field angle $\geq$30$^{\circ}$. The M87 jet strongly collimates  
at 30-100 Schwarzschild radii (r$_s$) from the BH, collimation continuing 
out to $\sim$1000 r$_s$. 2) The semicontinuous 
emanation of the jets in the infrared and radio during time 
intervals of the order of minutes, as observed in GRS 1915+105 
(Eikenberry et al. 1998; Fender \& Pooley, 1998; Mirabel et al. 1998). 

Although several numerical simulations have been made, 
there are still no definitive tests to discriminate the relative 
role of these two MHD mechanisms.

\vskip .05in
{\bf What are the QPOs of maximum fix frequency ?}

Because QPOs with maximum fix frequency have been observed many times 
in some microquasars (e.g. 67 Hz in GRS 1915+105; Morgan, Remillard \& 
Greiner, 1997), it is believed that 
they are related to fundamental properties of the BHs, such as its  mass 
and spin. The problem is that there are more than 3 different alternative 
explanations in terms of General Relativity (see Mirabel \& Rodr\'\i guez, 
1999 for references). The theories should provide 
the observational tests that would discriminate among the models proposed.  

\vskip .05in
{\bf Are the jets discrete plasmons or semi-continuous flows with internal and/or external shocks ?}

Internal shock models of microquasar jets (Bodo \& Ghisellini, 1995; 
Kaiser, Sunyaev \& Spruit, 2000) propose that the plasma 
velocity is smaller than the pattern velocity. Outbursts 
in the core lit up the jets are by shock fronts that travel along the 
jets which accelerate the 
relativistic particules that emit the synchrotron radiation. Internal 
shock models 
relax the requirements on the power of the central engine because much of 
the energy underlying the outbursts is stored in the continuous jet. 
An alternative model of the twin moving radio lobes observed in Sco X-1 
(Fomalont, Geldzahler \& Bradshaw, 2000) is that they are 
intetraction of the energy 
flow from beams with the interstellar medium. The lobes advance at 
$\sim$0.45c but the beam velocity is $\geq$0.95c. While the observations 
of GRS 1915+105 and SS433 would be more consistent with the internal shock 
model, the moving lobes in Sco X-1 seem to be shocks in working surfaces of  
an external medium. While the observed velocities of the ejecta 
in GRS 1915+105 ($\sim$ 0.95c) and SS 433 ($\sim$ 0.26c) have been the 
same over several years, whereas in Sco X-1 the speeds for different pairs of 
components at different times range between 0.31c and 0.57c. This suggests  
that different physical processes may dominate in different sources.     

Jets in microquasars, $\gamma$-ray burst (GRB) afterglows and AGN 
show analogous phenomenologies. 
It is interesting that the internal shock model originally developed 
for GRBs (M\'esz\'aros \& Rees 1997) and the external shock model originally 
proposed for the terminal lobes in AGNs (Blandford \& Rees, 1974) 
are now being applied to microquasars (Kaiser, Sunyaev \& Spruit 2000; 
Fomalont, Geldzahler \& Bradshaw, 2000). Conversely, the plasmon model 
originally proposed to interpret the microquasar jets is being proposed 
in the ``cannonball" model of GRBs by Dar \& De R\'ujula (2000). 

\vskip .05in
  
{\bf Have extragalactic microquasars been identified ?}

Ultraluminous X-ray compact sources  
have been identified in several nearby spiral galaxies (Colbert \& Mushotzky, 
1999; Makishima et al. 2000) as well as in dwarf galaxies 
(Mirioni \& Pakull, 2000). These enigmatic sources have luminosities 
in the range of L$_x$ = 10$^{38-40}$ erg s$^{-1}$, and it has been 
proposed by Colbert \& Mushotzky (1999) that they are BHs of 10$^{2-4}$ M$_{\odot}$. Makishima et al. (2000) 
propose that some of these superluminous sources are BHs   
as the microquasars GRS 1915+105 and GRO J1655-40, with masses 
below $\sim$100 M$_{\odot}$ and high 
disk temperatures  because rapid rotation gets the disk closer to the BH, 
hence hotter. Unfortunately, the VLBA array does not have the sensitivity 
to image in its plateau state extragalactic ``super-microquasars" as GRS 1915+105. 
\vskip .05in

{\bf Why superluminal microquasars have unbroken power law photon spectra ?}

Grove (1999) has shown that the microquasars with superluminal jets 
have power-law-gamma-ray states with an index of 2.5-3. Contrary to 
Cygnus X-1 and 1E1740.7-2942, the superluminal sources 
GRO J1655-40 and GRS 1915+105 have 
unbroken power laws up to almost 1 MeV with no indication of Comptonization. 
A similar photon spectrum has been observed by SAX in XTE 1118+48 (Frontera 
et al. 2001). Since we don't know 
what is the origin of the electrons that produce the gamma-ray photons it
is still unclear why some BHs would have broken, whereas others have 
unbroken power law photon spectra. The sensitivity of INTEGRAL in this 
domain of energy will certainly provide new perspectives on this issue.  

\vskip .05in

{\bf Will microblazars be found ?} 

In most microquasars where $\theta$ (the angle 
between the line of sight and the axis of ejection) has been determined,
large values are found ($\theta \geq 70^\circ$). Namely, except some 
exceptions (as in the case of Sco X-1 where 
$\theta$ = 45$^\circ$; Fomalont, Geldzahler \& 
Bradshaw, 2000), the axis of ejection in most cases is close to the 
plane of the sky. This is consistent 
with the statistical expectation since the probability of
finding a source with a given $\theta$ 
is proportional to $sin~\theta$. We then expect to find as many
objects in the $60^\circ \leq \theta \leq 90^\circ$ range
as in the $0^\circ \leq \theta \leq 60^\circ$ range.
However, this argument suggests that we should eventually detect
objects with a small $\theta$. For objects with $\theta \leq 10^\circ$
we expect the timescales to be shortened by
0.5$\gamma$$^2$ and the flux densities to be boosted by 8$\gamma^3$ 
with respect to  
the values in the rest frame of the condensation.
For instance, for motions with $v$ = 0.98c ($\gamma$ = 5), the timescale will shorten
by a factor of $\sim$50, the flux densities will be boosted by
a factor of $\sim 10^3$, and the photon spectrum of the source will be very hard. Then, for a galactic source with
relativistic jets and small $\theta$ we expect fast and intense
variations in the observed flux.
Microblazars may be quite hard to detect
in practice, both because of the low probability
of small $\theta$ values and because of the fast decline
in the flux.

\vskip .05in

{\bf Could microquasars be unidentified sources of 
gamma-rays detected by EGRET ?}

There is the indication that the synchrotron component of the jets 
in microquasars could reach the X-ray domain (Markoff, Falcke \& Fender, 
2000). This would require the ejection of plasma containing electrons 
with Lorentz factors $\geq$10$^7$, as in Blazars. In this context, the recent 
observation by Paredes 
et al. (2000) of persistent radio jets from LS5039, which is located in the 
error box of an EGRET source raises the possibility that microquasars 
could be persistent sources of gamma-rays. 

\vskip .05in

{\bf Could microquasars be sources of $\geq$10$^{20}$ eV cosmic rays ?}

Due to opacity by cosmic infrared photons, the sources of $\sim$10$^{20}$ 
eV cosmic rays must be within 50 Mpc from the Galaxy. AGNs have been proposed 
as possible sources of such high energy particles. As shown above, the bulk 
and intrinsic 
Lorentz factors of the electrons in microquasars are comparable to 
those in AGNs. Since they are much closer and numerous than the later, 
the very high energy cosmic rays could be produced by shocks in 
microquasar jets.

\vskip .05in

{\bf Are Gamma-ray-bursts extreme microquasars ?}

Gamma-ray bursts are at cosmological distances and ultra-relativistic 
bulk motion and beaming appear as essential ingredients to 
solve the enormous energy requirements (Castro-Tirado et al. 1999). 
Beaming reduces 
the energy release by the beaming factor f = $\Delta$$\Omega$/4$\pi$, where 
$\Delta$$\Omega$ is the solid angle of the beamed emission. 
Additionally, the photon energies can be boosted to higher
values.
BHs formed by core collapse producing outflows with bulk 
Lorentz factors $>$ 100 have been proposed 
as sources of $\gamma$-ray bursts (M\'esz\'aros \& Rees 1997).
Recent studies of gamma-ray afterglows suggest that they are highly 
collimated jets since breaks and a steepening from a power law in 
time t proportional to 
t$^{-1.2}$, ultimately approaching 
a slope t$^{-2.5}$ have been observed in light curves 
(Castro-Tirado et al. 1999). 

It is interesting that 
the power laws that 
describe the light curves of the ejecta in microquasars 
show similar breaks and steepening of the 
radio flux density (Rodr\'\i guez \& Mirabel, 1999). 
In microquasars, these breaks and steepenings have been 
interpreted (Hjellming \& 
Johnston 1988) as a transition from slow intrinsic expansion to 
free expansion in two dimensions. Besides, linear polarizations of 
about 2\% were recently measured in the optical afterglows, 
providing strong evidence that the afterglow radiation from gamma-ray 
bursters is, at least in part, produced by synchrotron processes. 
Linear polarizations in the range of 2-10\% have also been measured  
in microquasars at radio (Rodr\'\i guez et al. 1995; 
Hannikainen et al. 2000), and optical (Scaltriti et al. 1997) wavelengths. 

 In this context, microquasars in our own Galaxy seem to be less extreme 
local analogs of the super-relativistic jets associated to the more 
distant $\gamma$-ray bursters, which are jets with Lorentz factors 2 
orders of magnitude larger. Therefore, the physical mechanism that 
launches the jets in $\gamma$-ray bursters are likely to be different 
to the microquasars. Furthermore, GRBs do not repeat and 
seem to be related to catastrophic 
events, and have much larger super-Eddington luminosities. Therefore, 
the scaling laws in terms of the black hole mass that are valid in the 
analogy between microquasars and quasars do not seem to apply in the case of $\gamma$-ray bursters.

\acknowledgements
This work was partially supported by Consejo Nacional de Investigaciones 
Cient\'\i ficas y T\'ecnicas de Argentina.

% The endnotes section will be placed here.

\end{article}

\begin{thebibliography}{}

\bibitem[]{} Belloni, T, M\'endez, M, King, AR, van der Klis, M,
van Paradijs, J. 1997, {\it Ap. J.} 479: L145-48 

\bibitem[]{} Blandford, RD, Payne, DG. 1982, {\it MNRAS} 199: 883

\bibitem[]{} Blandford, RD \& Rees, M.J. 1974, MNRAS 169, 395

\bibitem[]{} Blandford, RD, Znajek, RL. 1977, {\it MNRAS} 179: 433 

\bibitem[]{} Bodo, G, Ghisellini, G. 1995, {\it Ap. J.} 441: L69-71

\bibitem[]{} Castro-Tirado, AJ. et al. 1994, Astrophys. J. Supp. Ser. 92, 469

\bibitem[]{} Castro-Tirado, AJ. et al. 1999, {\it Science} 283: 2069-73

\bibitem[]{} Colbert, E.J.M. \& Mushotzky, R.F. 1999, ApJ 519, 89

\bibitem[]{} Dar, A. \& De R\'ujula, A. 2000, astro-ph/0008474

\bibitem[]{} Dhawan, V., Mirabel, I.F. \& Rodr\'\i guez, L.F. 2000, ApJ 543, 
373

\bibitem[]{} Eikenberry, SS, Matthews, K, Morgan, EH, Remillard, RA, Nelson, RW.
1998, {\it Ap. J.} 494: L61-64

\bibitem[]{} Fender, R.P. 2000, astro-ph/0008447

\bibitem[]{} Fender, RP, Pooley, GG. 1998, {\it MNRAS} 300: 573-76

\bibitem[]{} Fomalont, E.B., Geldzahler, B.J. \& Bradshaw, C.F. 2000, 
submitted to ApJ Letters

\bibitem[]{} Frontera, L. et al. 2000, Proceedings of the IV Integral meeting, 
Alicante, Sept. 2000

\bibitem[]{} Greiner, J., Morgan, E.H. \& Remillard, R.A. 1996, ApJ 473, L107

\bibitem[]{} Grove, J.E. 1999, ASP Conference Series, 161, p54


\bibitem[]{} Hannikainen, DC, et al. 2000, ApJ 540, 521


\bibitem[]{} Hjellming, RM, Johnston, KJ. 1988,
{\it Ap. J.} 328: 600-09


\bibitem[]{} Junor, W., Birettta, J.A. \& Livio, M. 1999, Nature 401, 891

\bibitem[]{} Kaiser, C.R., Sunyaev, R. \& Spruit, H.C. 2000, A\&A 356, 975 

\bibitem[]{} Livio, M. 1999, Physics Reports 311, 225

\bibitem[]{} Makishima, K. et al. 2000, ApJ 535, 632

\bibitem[]{} Markoff, S., Falcke, H. \& Fender, R. 2000, astro-ph/0010560

\bibitem[]{} Meier, D.L, Edgington, S, Godon, P, Payne, DG, 
Lind, KR. 1997, {\it Nature} 388: 350-52

\bibitem[]{} Meier, D.L. 2001, Review in this issue.

\bibitem[]{} M\'esz\'aros, P, Rees, MJ. 1997,
{\it Ap. J.} 482: L29-32

\bibitem[]{} Mirabel, I.F. \& Rodr\'\i guez, L.F. 1998, Nature, 392, 673

\bibitem[]{} Mirabel, I.F. \& Rodr\'\i guez, L.F. 1999, ARAA 37, 409

\bibitem[]{} Mirabel, I.F. et al. 1998, A\&A 330, L9

\bibitem[]{} Mirioni, L. \& Pakul, M. 2000, Private communication

\bibitem[]{} Morgan, E.H., Remillard, R.A. \& Greiner, J. 1997, ApJ 482, 993

\bibitem[]{} Paredes, J. M., Marti, J., Rib\'o, M., Massi, M. 2000, Science 
288, 2340

\bibitem[]{} Rodr\'\i guez, L.F. et al. 1995, Ap.J.Supp. 101, 173

\bibitem[]{} Rodr\'\i guez, LF, Mirabel, IF. 1999, {\it Astron. Astrophys.} 
340: L47-50

\bibitem[]{} Sams, B.J., Eckart, A. \& Sunyaev, R. 1996, Nature 382, 47

\bibitem[]{} Scaltriti, F. et al. 1997, A\&A 325, L29

\bibitem[]{} Zhang, N.S., Cui, W. \& Chen, W. 1997, ApJ 482, L155


\end{thebibliography}
\end{document}